\documentclass[twocolumn,10pt]{article} 

\usepackage[square,numbers,sort&compress,comma]{natbib}
\usepackage{xcolor}
\usepackage{amsmath}
\usepackage{amssymb}
\usepackage{caption}
\usepackage{graphicx}
\usepackage{latexsym}
\usepackage{times}
\usepackage{comment}

\topmargin - 12pt 
\renewenvironment{abstract}%
              {
               \small
               {\bfseries \abstractname}
               \par
               \vspace{10pt}
              }

\renewcommand\abstractname{Abstract}

\newcommand{\nomenclature}
              [1]
              {
               \bgroup
               \flushleft
               \small\bf
               #1
               \par
               \egroup
              }

\renewcommand{\section}
              [1]
              {
               \bgroup
               \flushleft
               \small\bf
               \refstepcounter{section}
               \arabic{section}. #1
               \par
               \egroup
              }

\renewcommand{\subsection}
              [1]
              {
               \bgroup
               \flushleft
               \small\em
               \refstepcounter{subsection}
               \arabic{section}.
               \arabic{subsection}. #1
               \par
               \egroup
              }

\renewcommand{\subsubsection}
              [1]
              {
               \bgroup
               \flushleft
               \small\em
               \refstepcounter{subsubsection}
               \arabic{section}.
               \arabic{subsection}.
               \arabic{subsubsection}. #1
               \par
               \egroup
              }

  \newcommand{\acknowledgement}
              [1]
              {
               \bgroup
               \flushleft
               \small\bf
               #1
               \par
               \egroup
              }

  \newcommand{\sectionbib}
              [1]
              {
               \bgroup
               \flushleft
               \small\bf
               #1
               \par
               \egroup
              }

\begin{document}

\title{\LARGE Turbulent flame speed of thermodiffusively unstable flames: experimental investigation and scaling} 

\author{{\large Guido Troiani$^{a,*}$, Pasquale Eduardo Lapenna$^{b}$, Francesco D'Alessio$^{b}$, Francesco Creta$^{b}$}\\[10pt]
        {\footnotesize \em $^a$Laboratory of Processes \& Systems Engineering for Energy Decarbonisation, ENEA Rome, Italy}\\[-5pt]
        {\footnotesize \em $^b$Department of Mechanical and Aerospace Engineering, Sapienza University of Rome, Italy}\\[-5pt]
        }

\date{}


\small
\baselineskip 10pt


\twocolumn[\begin{@twocolumnfalse}
\vspace{50pt}
\maketitle
\vspace{40pt}
\rule{\textwidth}{0.5pt}
\begin{abstract} 
This work presents an experimental set of Bunsen flames characterized by a moderate Reynolds number and a variable turbulence intensity. Ten lean hydrogen-enriched methane-air mixtures at three levels of turbulence are investigated, ranging from pure methane-air to pure hydrogen-air. Such mixtures are selected in order to have an almost constant laminar flame speed while inducing the onset of thermal-diffusive (TD) instability by gradually increasing the hydrogen content of the blend. The flames are analyzed in terms of global consumption speed, stretch factor, and flame surface area. Results indicate an interplay between TD instability and turbulence that enhances the overall flame propagation. In particular, below a transitional Lewis number, flame propagation is observed to be particularly sensitive to external turbulent forcing, expressed in terms of the Karlovitz number.
A parameterization is thus proposed, based on a functional form depending on both Karlovitz and Lewis numbers, able to fit the experimental results at different turbulence levels and capture the steep transition across the transitional Lewis number.
%
\end{abstract}
\vspace{10pt}
\parbox{1.0\textwidth}{\footnotesize {\em Keywords:} Hydrogen; Turbulent combustion speed; Thermal-diffusive instability; Lewis number effects; Preferential diffusion; Turbulent combustion parameterization: Experimental analysis; Spectroscopy;}
\rule{\textwidth}{0.5pt}
\vspace{10pt}

\end{@twocolumnfalse}] 

\clearpage


\section{Introduction\label{sec:introduction}} \addvspace{10pt}
In the context of turbulent premixed combustion, the turbulent combustion speed $S_T$, defined as the speed at which 
reactants are consumed through a suitably defined average flame surface, is one of the fundamental quantities to be modeled.
An initial attempt to parameterize this quantity dates back to the seminal work of Damkh\"oler \cite{damkohler1940einfluss},   
and the search for a model of the turbulent combustion speed has been relentless ever since \cite{abdel1985lewis,haworth1992numerical,law2004effects,muppala2005development,lipatnikov2023experimental}.

For carbon-based fuels, where thermal and mass diffusivities can generally be of the same order at near-stoichiometric conditions, the dependence of turbulent combustion speed is sought in terms of observables that can affect the amount of flame wrinkling, such as the whole spectrum of turbulent kinetic energy and its dissipation rate, usually expressed in terms of the turbulent velocity fluctuations, integral and Taylor scales. The flame's thermochemical parameters also play their role, namely thermal expansion, thermal thickness, laminar flame speed, whether strained or unstrained. Usually, such quantities are used in non-dimensional form, i.e., the turbulent velocity fluctuations in units of laminar flame speed $u'/S_L$, thermal thickness in units of integral length-scale $\delta_{th}/\ell_0$, the Schmidt or Zel'dovich numbers as well as the Markstein number.    
In this context, much work was done to model the experimental measurements of the turbulent combustion speed \cite{kolla2009scalar,peters1999turbulent,gulder1991turbulent,zimont2000gas,bradley1992fast}. 

Following a slightly different approach, attempts were also made to decouple the effects of the turbulent area $A_T$ from the diffusive effects acting within the flame front thickness. To this end, the turbulent combustion speed was modeled as $S_T/S_L^0 \sim I_0 (A_T/A_0)$, where $A_0$ is the unwrinkled mean flame area and the stretching factor $I_0$ is a measure of the discrepancy between the increase of turbulent combustion speed and the increase of turbulent area~\cite{bell2002numerical,hawkes2006comparison,troiani2013turbulent,lapenna2021mitigation}, effectively accounting for the variation of flame reactivity due to turbulence.
An interesting scaling for $I_0$ with the Karlovitz number was proposed in \cite{nivarti2019reconciling}, namely $I_0 \sim Ka^{(1/3)}$, which well fits the experimental results of \cite{wabel2017turbulent,yuen2013turbulent}
in which, again, the fuel Lewis number is kept close to one. 


An additional effect to be taken into account is the onset of the intrinsic hydrodynamic or Darrieus-Landau (DL) instability, which can potentially affect propagation at large scales. Asymptotic theory and more recently Direct Numerical simulations (DNS) \cite{matalon1982flames,clavin1982effects,al2019darrieus,frouzakis2015numerical,berger2023flame} have performed linear stability analyses for premixed flames yielding dispersion relations for the growth rate of harmonic flame disturbances. A cut-off lengthscale $\lambda_c$ emerges so that any disturbance larger than this is exponentially amplified in the linear regime. In practical situations, when the integral scale of the system is larger than the cut-off scale, the flame can exhibit a range of unstable scales that further enhance the wrinkling \cite{creta2016interplay,lapenna2019large,troiani2015experimental,rocco2015curvature}. although at higher turbulence intensity a unified turbulence-dominated regime is likely reached  \cite{yang2018role,lapenna2021mitigation,troiani2022self}.

\begin{figure}[h!]
\centering
\includegraphics[width=1 \columnwidth]{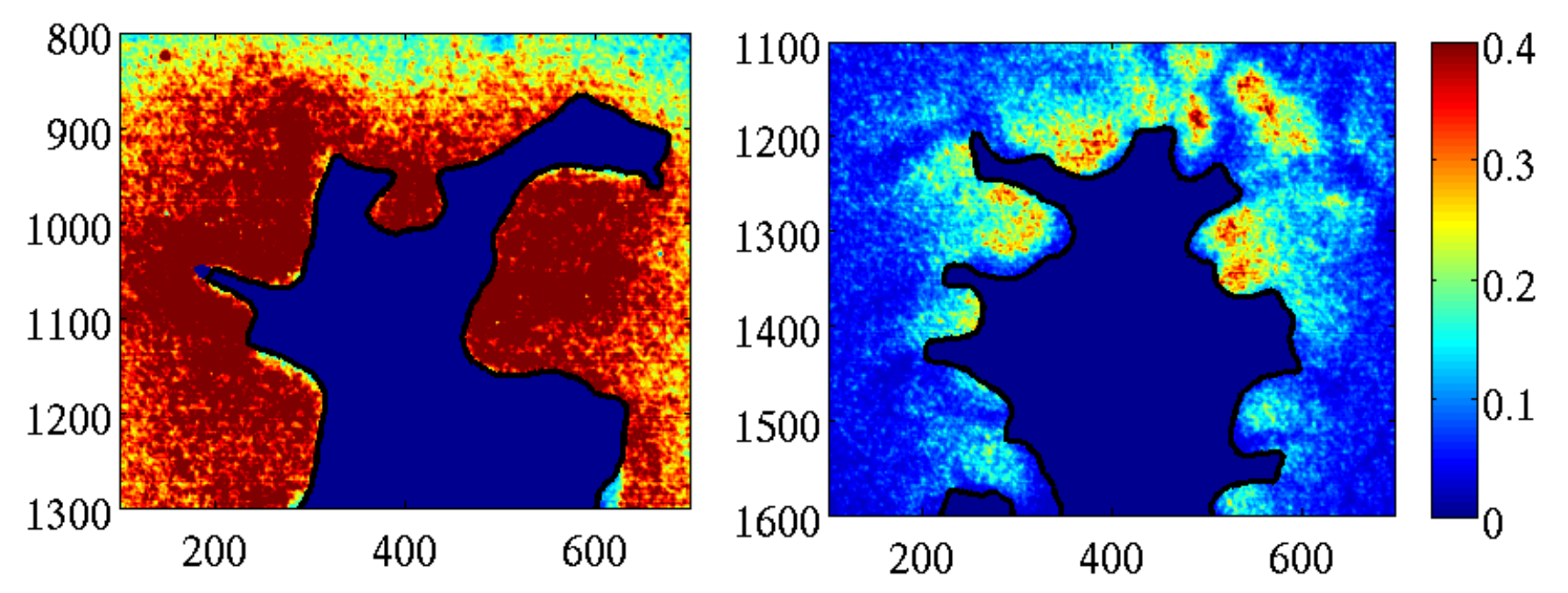}
\caption{OH-LIF measurements for lean methane-air flame (left panel) and lean hydrogen-air flame (right panel), 
as reported in Table~\ref{tab:thermochemistry}. LIF signal is normalized with local maximum of each snapshot. In pure hydrogen flame it is visible the enhancement and depletion of  LIF intensity signal that can be translated into and indication of the intensity of the reaction rate.  Resolution is about 50 $\rm{\mu m}$ per pixel. }
\label{fig:LIF_snap}
\end{figure}

The use of hydrogenated carbon-based fuels to mitigate carbon dioxide emissions introduced the additional effect of uneven heat and fuel mass diffusivities, resulting in Lewis numbers potentially significantly smaller than unity. 
This leads to the activation of an additional intrinsic instability mechanism, of thermodiffusive (TD) nature and active at smaller scale, which has been observed, both numerically and experimentally, to deeply affect both laminar and turbulent flame propagation \cite{frouzakis2015numerical,creta2020propagation,berger2022synergistic,lipatnikov2023experimental,Troiani_ECM2023}.
Figure \ref{fig:LIF_snap} shows OH-LIF measurements performed by the authors, used to trace the reactivity and morphology of lean methane-air and hydrogen-air Bunsen flames. On one hand TD instabilities, which introduce a distinct cellularity in the flame conformation, play a distinct role in increasing the turbulent area. On the other, the induced stretch patterns locally increase the reaction rate, affecting the flame reactivity \cite{rocco2015curvature,berger2022synergistic}. The two effects combined, result in an enhanced combustion regime typical of low Lewis number flames.


In this work we present results from an experimental campaign in which the addition of hydrogen in a methane-air Bunsen flame at constant bulk Reynolds number (mixtures ranging from pure methane to pure hydrogen) drives the Lewis number from almost unity toward lower values. 
Such a set of variable $Le$ number flames is forced with different levels of turbulence, achieved by moving a perforated plate along the streamwise direction. Mixtures are carefully chosen so as to maintain the laminar flame speed constant while the Lewis number is decreased, so as to isolate the Le effects and the eventual synergistic effects with turbulence. Results show that the enhancement of the turbulent combustion speed and its sensitivity to the Karlovitz number, are subject to a significant change at a transitional or critical Lewis  number, which can be attributed to the onset of TD-instability. Previous turbulent combustion speed models where the influence of the Lewis number is modelled with a single exponent, such as \cite{bradley1992fast}, do not seem to capture such transition.
For this reason, a parameterization is proposed, based on a functional form depending on both Karlovitz and Lewis numbers, capturing the transition across the critical Le number for all turbulence levels.



\section{Methods\label{sec:Methods}} \addvspace{10pt}

\subsection{Thermochemical properties\label{subsec:subsection1}} \addvspace{10pt}

The developed experimental dataset features 10 different fuel-air mixtures at increasing content of hydrogen, ranging from pure CH$_4$ to pure H$_2$. Each condition is fully characterized by the equivalence ratio $\phi$ and by the volumetric fraction of hydrogen $\alpha$ in the fuel. The $(\phi,\alpha)$ combinations for the present analysis are chosen such that the unstretched laminar flame speed $S^0_L$ remain approximately constant in order to not introduce any bias in the turbulence-flame interactions interpretation. As the hydrogen content is increased, differential diffusion effects are expected to be more relevant. Such effects are usually quantified by means of the definition of the Lewis number for the fuel. However, when fuel blends and realistic mixtures are considered, as in the case of hydrogenated methane-air mixtures, multiple Lewis numbers can be singled out and the definition of an effective Lewis number $Le$ can be useful. Such effective value is evaluated using the definition of Bechtold et al.~\cite{bechtold2001dependence} using a single reference Lewis number for the fuels mixture with a volume-based approach~\cite{bouvet2013effective}. The oxidizer and fuels Lewis numbers are taken on the burned side of a laminar 1D premixed flame, simulated using Cantera~\cite{cantera} with detailed chemistry and transport models, at each condition of interest in terms of $\phi$ and $\alpha$. Using the same set of 1D flames the Zel'dovich number $Ze$ is evaluated as recently described in~\cite{berger2022intrinsic1}. The resulting $Le$ ranges between $1.03$ and $0.38$ as $\alpha$ is increased for the 10 thermochemical conditions used in this study. Note that different values of $Le$ can be obtained using different definitions for $Ze$ and the Lewis numbers, however a maximum difference of $\sim5\%$ has been obtained using alternative definitions. The values of $Le$ and the laminar unstretched flame speed $S_L^0$ and laminar thermal flame thickness $\ell_T$ are reported in Tab.~\ref{tab:thermochemistry}.

\begin{table}[h!]
  \begin{center}
\footnotesize
    \caption{Thermochemical conditions of the experimental dataset: equivalence ratio $\phi$, volumetric fraction of hydrogen in the fuel $\alpha$, thermal flame thickness $\ell_T$, laminar flame speed $S^0_L$ and effective Lewis number $Le$. }
    \label{tab:thermochemistry}
    \begin{tabular}{cccccc} 
          \hline
         case\# & $\phi$~[-]  & $\alpha$~[-]  & $\ell_T$~[mm] & $S^0_L$~[m/s] & $Le$\\
      
            \hline
      1 & 0.875 & 0.0  & 0.48 & 0.33 & 1.03 \\
      2 & 0.825 & 0.3  & 0.47 & 0.34 & 0.92 \\
      3 & 0.775 & 0.3  & 0.49 & 0.32 & 0.85 \\
      4 & 0.725 & 0.46 & 0.48 & 0.32 & 0.76 \\
      5 & 0.675 & 0.6  & 0.48 & 0.33 & 0.67 \\     
      6 & 0.650 & 0.66 & 0.49 & 0.32 & 0.62 \\     
      7 & 0.625 & 0.72 & 0.49 & 0.32 & 0.59 \\     
      8 & 0.600 & 0.78 & 0.49 & 0.33 & 0.54 \\     
      9 & 0.550 & 0.88 & 0.49 & 0.33 & 0.47 \\     
     10 & 0.475 & 1.0  & 0.51 & 0.33 & 0.38 \\     
            \hline
    \end{tabular}
  \end{center}
\end{table}

Once the thermochemical parameters have been evaluated, it is possible to estimate the onset of TD-instability by means of the definition of a critical Lewis number $Le_0$. When the mixture $Le$ is smaller $Le_0$, TD effects are destabilizing a perturbed planar flame while, in the opposite case only the ubiquitous DL effects remain active to destabilize the flame. Possible  definitions of $Le_0$ stems from model dispersion relations, identifying the Lewis values that lead to a null second order term in the dispersion relation as discussed in~\cite{creta2020propagation}. In the present case, using realistic mixtures, we employ the model dispersion relation of Matalon et al.~\cite{matalon2003hydrodynamic} as also done in recent DNS studies~\cite{berger2022intrinsic1,howarth2022empirical}. However, it is worth mentioning that the resulting $Le_0$ is only a qualitative indicator, as the model dispersion relation can largely differ from the dispersion relations obtained using DNS~\cite{berger2022intrinsic1,lapenna2021subgrid}. This being said, the resulting $Le_0$ remains in a rather confined range between $0.50$ and $0.63$, confirming that the thermochemical flame conditions investigated in this work span both TD-stable (low $\alpha$) and TD-unstable mixtures (high $\alpha$).

\subsection{Turbulence characteristics and measurements \label{subsec:subsection2}} \addvspace{10pt}
The flames investigated in this work feature a Bunsen burner configuration with a diameter of $D = 18$~mm. The same geometry and set-up was employed in previous studies focusing on the interaction between DL instability and turbulence \cite{troiani2015experimental,creta2016interplay,lamioni2019strain,lapenna2021mitigation,troiani2022self} at different Reynolds numbers and propane-air mixtures. For the present study, a moderate bulk Reynolds number $Re\sim5000$ is chosen and the turbulence intensity is further modulated by a grid placed before the nozzle exit. In particular, three positions of the grid are chosen in order to reach different levels of axial velocity fluctuations $u'$, namely located at $1D$, $2.5D$ and $6D$ from the nozzle exit. As expected, higher $u'$ and larger turbulence integral scale $\ell_0$ are obtained when the grid is located closer to the nozzle exit as reported in Tab.~\ref{tab:turbulence}. 
\begin{table}[h!]
\centering 
\footnotesize
    \caption{Turbulent non-reacting flow field characterization of the Bunsen configuration at $Re\sim5000$ for the three different grid positions resulting in \textit{low}, \textit{medium} and \textit{high} $u'$.}
    \label{tab:turbulence}
    \begin{tabular}{cccc} 
    \hline
      case label: & \textit{low} & \textit{medium} & \textit{high}  \\
            \hline
      grid position & $6.0~D$ & $2.5~D$ & $1.0~D$  \\
      $u'$~[m/s] & $0.22$ & $0.33$ & $0.44$  \\
      $\ell_0$~[mm] & $5.5$ & $6.2$ & $6.9$  \\
      plots legend  & \textcolor{red}{$\bigcirc$}  & \textcolor{blue}{$\mathbf{\square}$} & $\mathbf{\triangle}$   \\
             \hline
    \end{tabular}
\end{table}
The axial velocity fluctuations are measured along the burner axis at its exit while the integral length scale is measured at the first zero crossing of the axial velocity autocorrelation coefficient along the burner axis.

The velocity fields are measured via particle image velocimetry (PIV), seeding $2~\mu {\rm m}$ alumina particles in the fresh mixture as a laser sheet scans an axial plane above the Bunsen exit. Mie diffusion of light resulting from the elastic interaction between the laser light and the alumina particles is recorded by a CCD camera with a repetition rate of $10$ Hz and a resolution of $2048 \times 2048$ pixels equipped with a $80$ mm focal length lens and an interference filter to suppress any other wavelength of light except that at $532$ nm, typical of $2^{nd}$ harmonic of Nd:YAG sources. The velocity field is obtained by $32 \times 32$ pixels interrogation windows with an overlap of $50\%$. The laser source is a $532$~nm Nd:YAG 
delivered through a cylindrical lens to expand the beam into a laser sheet characterized by a thickness of $\sim400$~$\mu$m at FWHM. The resolution for each pixel is $41$~$\mu$m. The resulting images are comprised of two distinct regions defined by very different levels of scattered light intensity reflecting unburnt and burnt regions with different values of temperature and density. The steep gradient separating the two regions traces the flame front separating burnt to unburnt mixture. A threshold technique allows for the definition of a binary image, where zero values are associated with cold denser reactants and ones stand for hot burned gases leading to the definition of a mean progress variable $\overline{C}$ ranging between $0$ and $1$. With binary images available, the flame front can be measured and values for the turbulent area as well as mean flame position follow. 

Radical concentration measurements are performed by acquiring the fluorescence signal emitted by hydroxyl (OH) radicals. To this end, a Nd:YAG laser beam combined with a tunable dye laser and with a second-harmonic generator crystal is used to shift the laser wavelength from $532$ nm down to $282.93$ nm (further experimental details can found in \cite{troiani2021fractal}). An intensified $2048 \times 2048$ pixels SCMOS camera, triggered with PIV CCD, collects the induced fluorescence emitted at $310$ nm.
The fluorescence signal from OH radical is proportional to its concentration and relative measurements are possible (absolute measurements of concentrations are instead prevented from non-radiative deexcitation channels that are generally active together with detectable fluorescence emission). 

\begin{figure}[h!]
\centering
\includegraphics[width=1 \columnwidth]{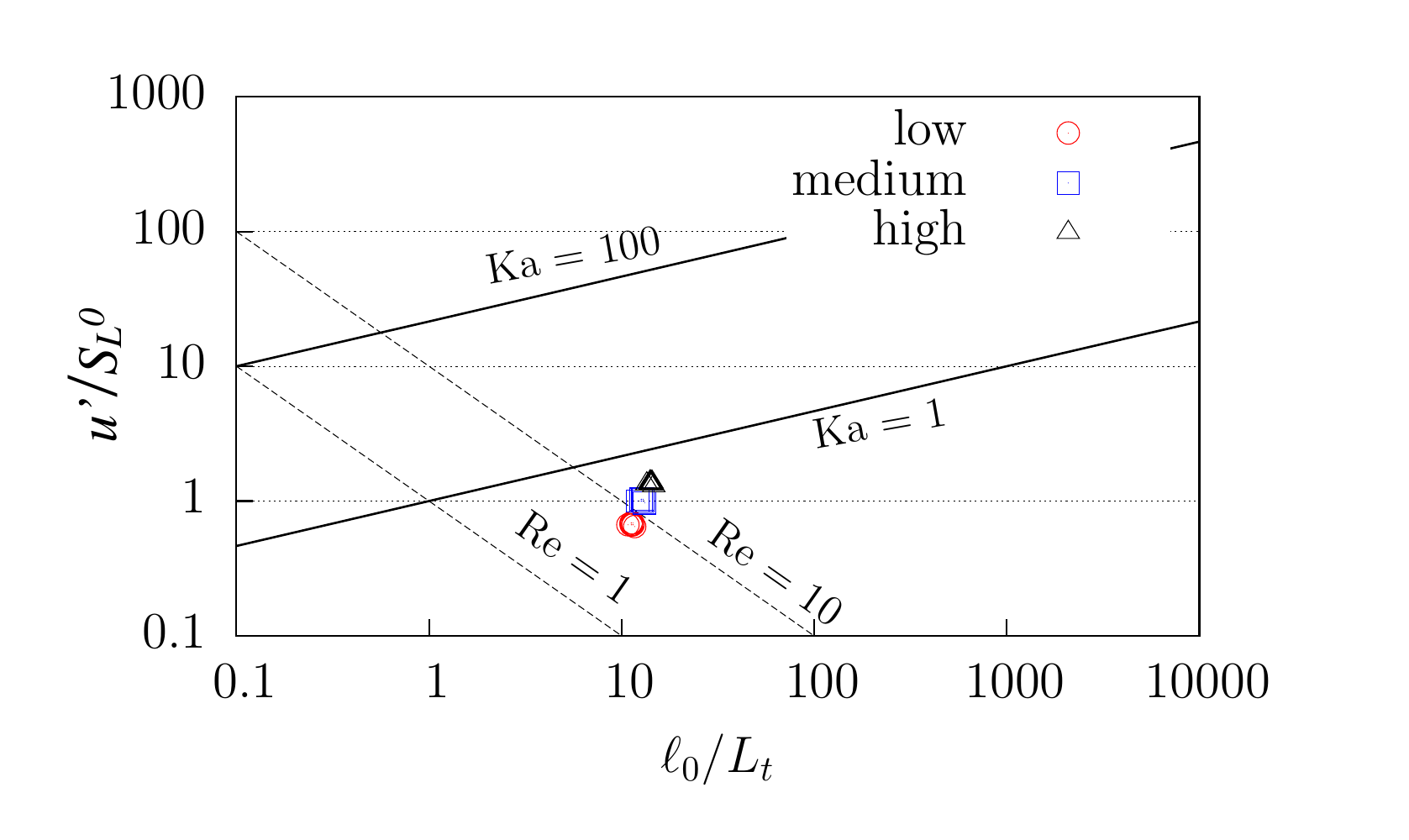}
\caption{Regime diagram for premixed turbulent combustion: each point represents the conditions of an experimental flame colored with the turbulence intensity which is varied by means of grid positioning as reported in Tables~\ref{tab:thermochemistry} and \ref{tab:turbulence}.}
\label{fig:Borghi}
\end{figure}

Since parameters $(\phi,\alpha)$ have been chosen for the present analysis such that  $S^0_L$ remains approximately constant,  the Karlovitz number $\text{Ka}= (u'/S^0_L)^{(3/2)} (\ell_T/\ell_0)^{(1/2)}$, where $\eta$ is he Kolmogorov scale, is also approximately constant. Indeed, the location of each flame series characterized by low, medium, and high values of $u'$ are rather close in the Borghi-Peters diagram as shown in Fig.~\ref{fig:Borghi}. 

\section{Results\label{sec:section3}} \addvspace{10pt}

The turbulent flame speed is evaluated using the global consumption speed concept~\cite{driscoll2008turbulent}:

\begin{equation}
    S_T=\frac{\dot{m}}{\rho_u A_0} 
    \label{S_T}
\end{equation}

\noindent where $\dot{m}$ is the inlet mass flow rate of the fresh mixture of density $\rho_u$, while $A_0$ represents the reference surface area of the mean flame, evaluated at the averaged progress variable $\overline{C} = 0.5$, when the time-averaged flame is considered as axisymmetric.

\begin{figure}[h!]
\centering
\includegraphics[width=0.95 \columnwidth]{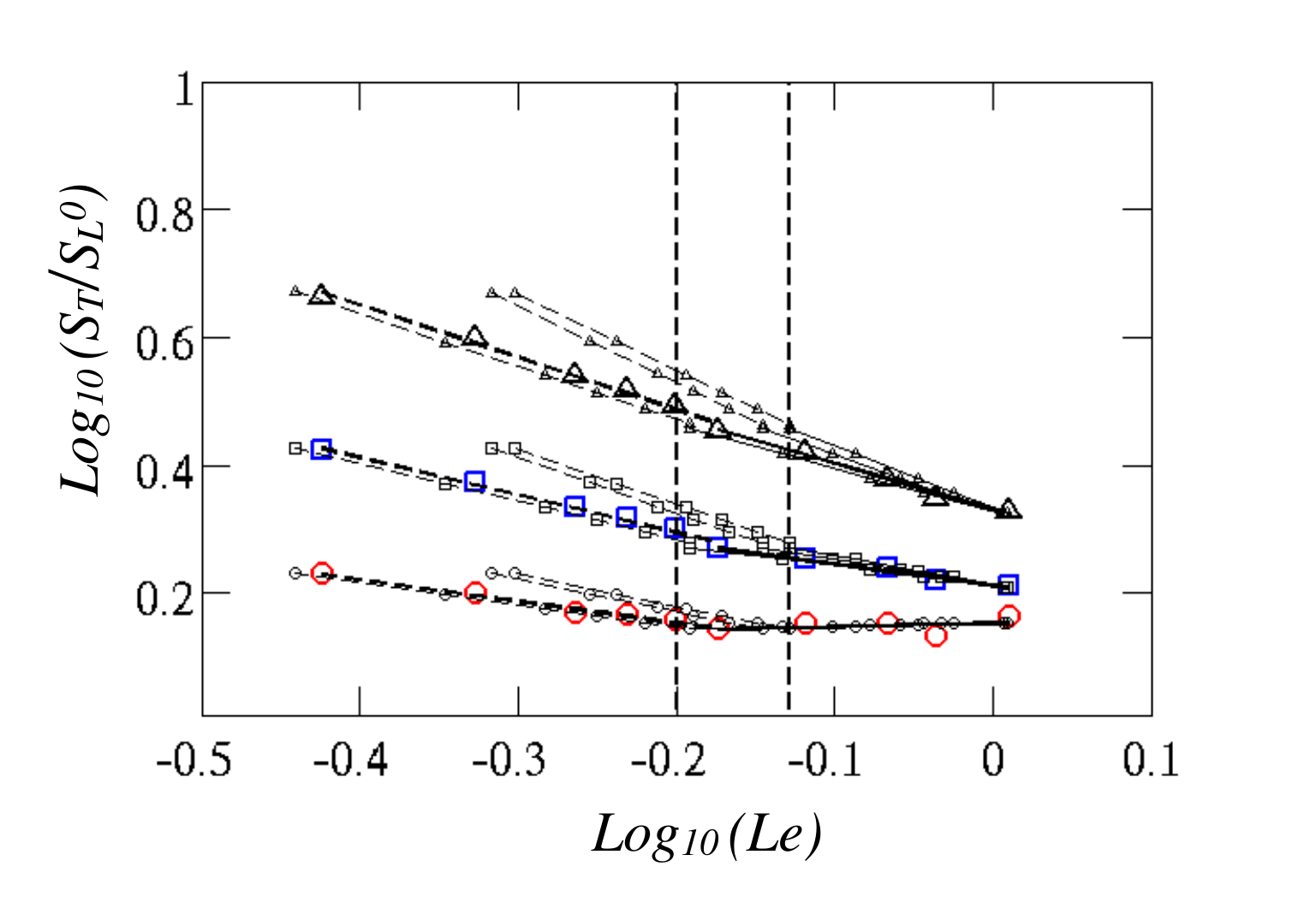}
\caption{\footnotesize Log-log plot of the turbulent combustion speed $S_T$ normalized with $S^0_L$. Dashed and continuous lines are linear regressions of experimental results conditioned to $Le<Le_0$ and $Le>Le_0$. For color coding, see figure \ref{fig:Borghi}.}
\label{fig:loglog_STSL}
\end{figure}

Figure~\ref{fig:loglog_STSL} shows the turbulent speed, normalized with the laminar unstrained speed $S_L^0$, as a function of $Le$ for the three turbulent conditions investigated. Large symbols refer to the $Le$ numbers reported in table~\ref{tab:thermochemistry}, while smaller symbols, used here as a reference, represent values of $Le$ obtained using alternative definitions, as discussed in the previous section. 
For the lower $u'$ case, the enhancement of $S_T/S_L^0$ is rather limited as the hydrogen content in the fuel blend is increased and the Lewis number lowered. Conversely, a substantial impact of the hydrogen addition can be observed for the medium $u'$ values and even more so for the large $u'$ cases. Comparing the effect of increasing turbulence at the same thermochemical conditions, i.e. same $Le$, it is evident that flames with higher hydrogen content are more sensitive to increasing $u'$. This effect can be preliminarily ascribed to the synergistic interaction of turbulence and thermal-diffusive flame instabilities described in a recent DNS study in \cite{berger2022synergistic}. In addition, the log-log visualization, not only reveals a linear scaling but also a visibly different behavior whether the effective Lewis number is larger or smaller than a transitional value. Such a value appears to depend on the particular approach used for the Lewis number evaluation but it is shown to be bounded within a rather narrow range comprised between $Le_ \simeq 0.63$ and $Le \simeq 0.74$, for all the three levels of turbulence. These limits, shown in Fig.~\ref{fig:loglog_STSL} by two vertical dashed lines, closely fall within the range of critical Lewis numbers $Le_0 = 0.5-0.63$ previously discussed. It is thus reasonable to assume a reference transitional Lewis number $Le_0^*$ between the two scaling behaviors across which the slope of $S_T/S^0_L$ is steepened as $Le < Le_0^*$. In addition, the transition between the two slopes is observed to be smoother when turbulence is increased. In other words, for flames with higher hydrogen content, the higher sensitivity to turbulence persists but the transition from thermal-diffusive unstable to stable regimes ceases to be abrupt. This behavior is akin to the unifying 'turbulence-dominated regime' observed for hydrodynamically stable and unstable flames \cite{lapenna2021mitigation,troiani2022self}.

To further investigate the global consumption speed, 
we express it as 
\begin{equation}
    \frac{S_T}{S_L^0} = I_0 \frac{A_T}{A_0} 
    \label{I_0}
\end{equation}
\noindent in which two contributions are singled out, namely
the turbulent surface area $A_T$, normalized with $A_0$, and the stretch factor $I_0$ which is a global measure of the deviation of the flame speed from that of an unstretched laminar flamelet. 
The turbulent area $A_T$ is determined from the volume integration of the flame surface density $\Sigma$ which is the ratio between the flame surface area and its embedding volume, as discussed in \cite{troiani2013turbulent,lamioni2019strain}, while the mean flame surface $A_0$ was defined in equation \ref{S_T}. Note that the choice of the reference surface $A_0$ becomes immaterial when equation \eqref{S_T} is plugged into equation \eqref{I_0}.

 \begin{figure}[h!]
\centering
\includegraphics[width=0.95 \columnwidth]{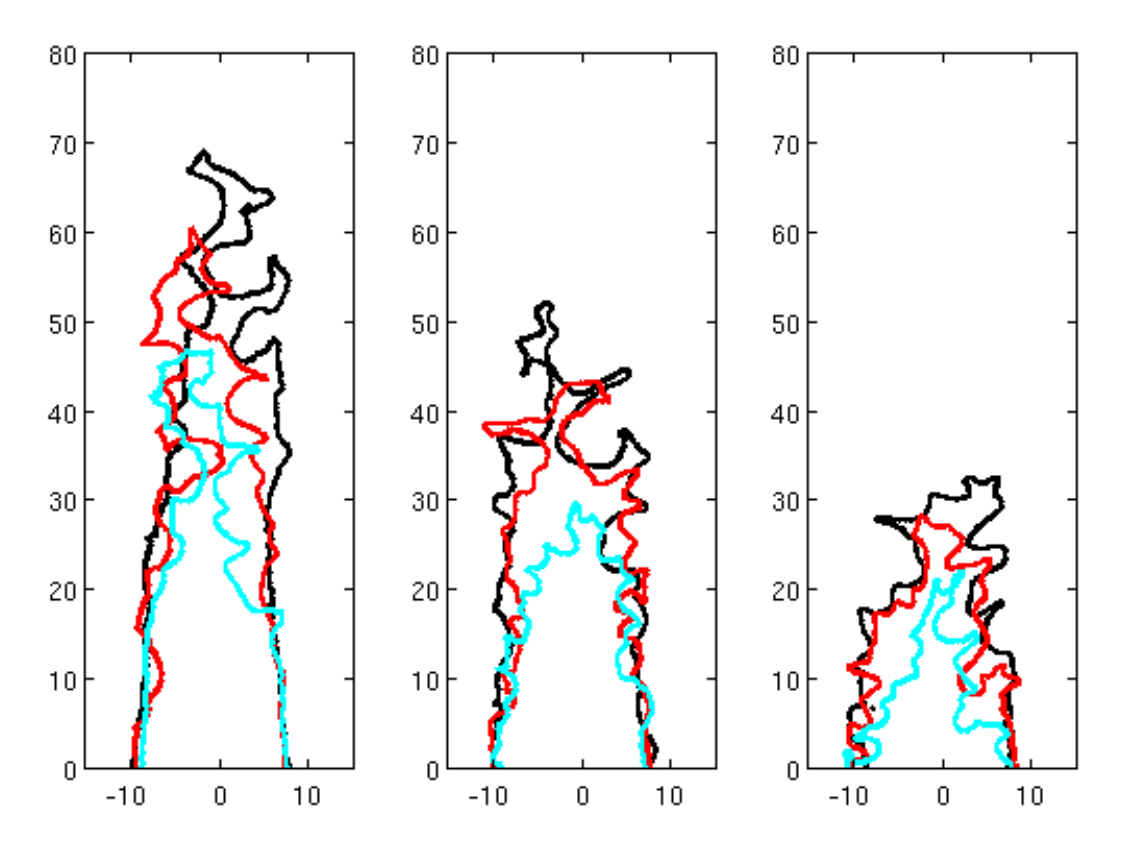}
\caption{\footnotesize Instantaneous flame fronts at different Lewis and Karlovitz numbers. Left, case \textit{low}; middle, case \textit{medium}; right, case \textit{high} (see Tab.\ref{tab:turbulence}). Colors: black, flame \#1; red, flame \#5; cyan, flame \#10 (see Tab.~\ref{tab:thermochemistry}). Axis dimensions,  mm.}
\label{fig:inst_flame}
\end{figure}

\begin{figure}[t!]
\centering
\includegraphics[width=0.95 \columnwidth]{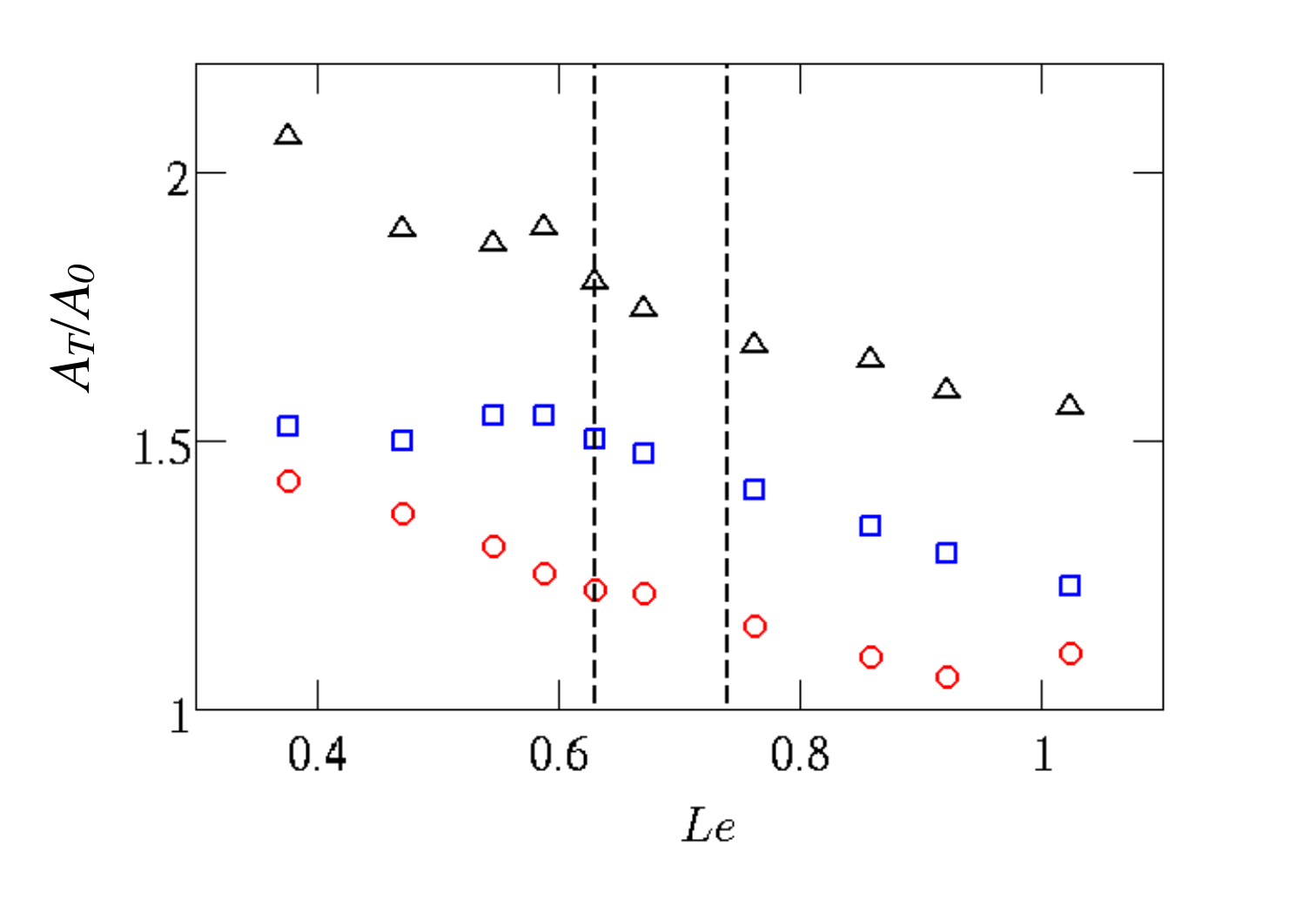}
\caption{\footnotesize  Behavior of the flame surface  area $A_T$ normalized the averaged flame surface area $A_0$ for different Lewis numbers and turbulence intensity. For color coding, see figure \ref{fig:Borghi}.}
\label{fig:ATA0}
\end{figure}
\begin{figure}[h!]
\centering
\includegraphics[width=0.97 \columnwidth]{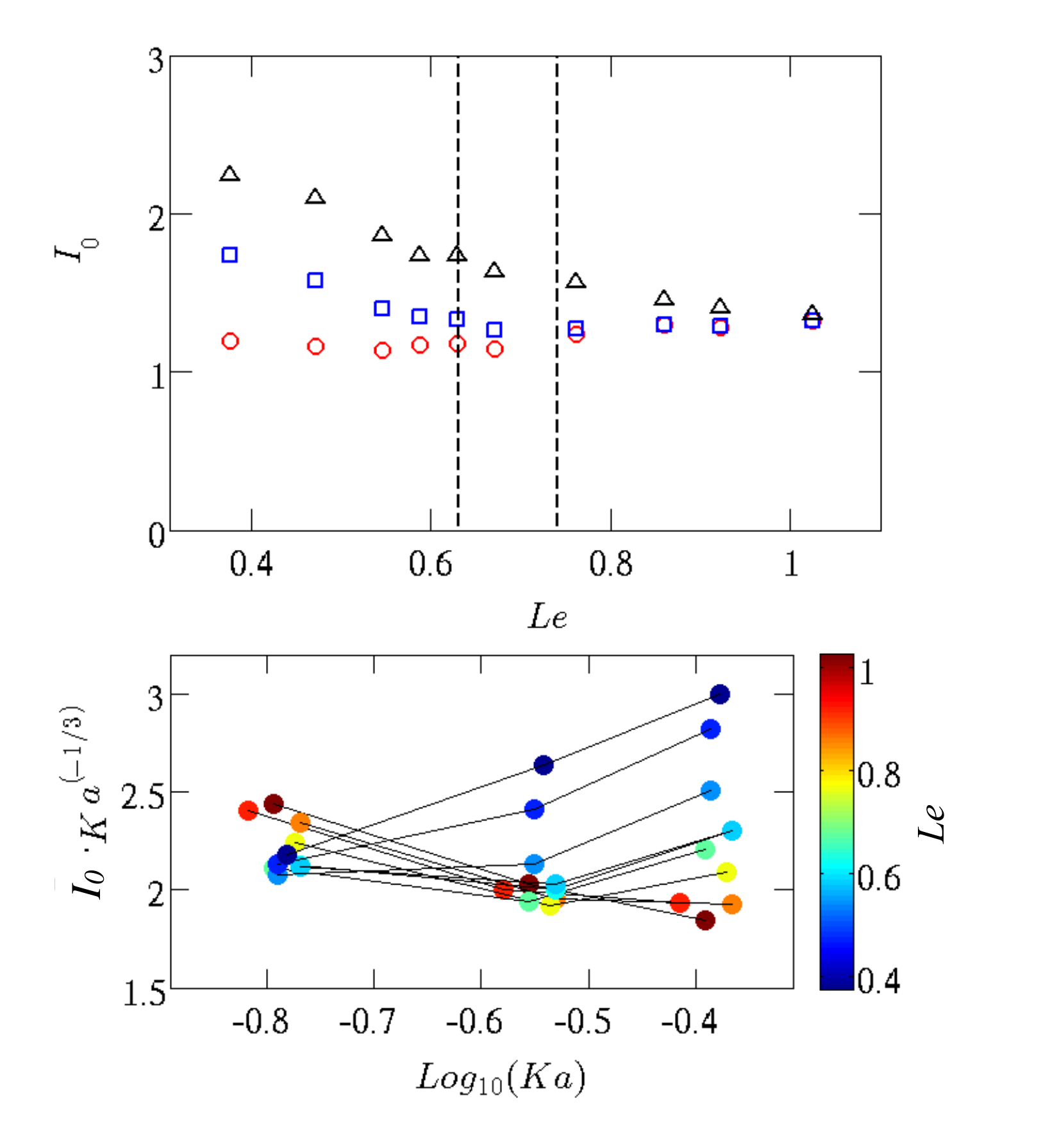}
\caption{\footnotesize Top: stretching factor $I_0$ at different operational conditions. Vertical dashed line refers to $Le_{eff} = 0.73$.  For color coding, see figure \ref{fig:Borghi}.
Bottom: stretching factor \textit{vs} Karlovitz number parameterized with $Le_{eff}$ reported in the colorbar. $I_0$ was compensated with $Ka^{(1/3)}$ according to \cite{nivarti2019reconciling}.}
\label{fig:I0}
\end{figure}

The instantaneous flame fronts shown in figure \ref{fig:inst_flame} underline the effects of both turbulence and Lewis number on the development of turbulent area. The increased amount of flame wrinkling acts to reduce the flame height as the turbulence is increased and the Lewis number is reduced.
A quantitative visualization of the normalized turbulent area $A_T/A_0$, which deprives the turbulent area of flame height effects and of the influence of the mean flame, is displayed in figure \ref{fig:ATA0} as a function of the Lewis number.
Although both thermal-diffusive effects and turbulence come into play in the enhancement of the normalized turbulent area, the trend appears to have a constant slope, insensitive to the turbulence intensity and Lewis number, and no transitional behavior is observed at the critical Lewis number $Le_0$. On the other hand, if we observe the stretching factor $I_0$, displayed in the top panel of figure \ref{fig:I0}, a clear transition emerges for $Le < Le_0$, wherein thermal-diffusive effects act synergistically with turbulence in amplifying $I_0$ and thus the net flame reactivity. This reveals that the transitional behavior shown in fig. \ref{fig:loglog_STSL} cannot be entirely accounted for by the flame surface area enhancement $A_T/A_0$, but should be sought mostly in the enhancement of the reactivity given by the synergistic action of turbulence (induced flame curvature and strain) and thermal-diffusive effects within the inner structure of the flame front, as revealed by the behavior of $I_0$.

The effect of the $Ka$ number at low Lewis numbers is more evident if the stretching factor is compensated with a scaling law of the kind $Ka^{1/3}$ (lower panel of figure \ref{fig:I0}) as discussed in \cite{nivarti2019reconciling}, for flames with thermal and mass diffusivity of the same order. In this latter scaling, the discrepancy between turbulent combustion enhancement $S_T/S_L^0$ and turbulent area increase $A_T/A_0$ is entirely accounted for by Karlovitz effects. For Lewis numbers closer to unity, the compensation yields a rather flat response ($I_0 \cdot Ka^{(-1/3)} \simeq const$), but as the $Le$ is further reduced and the thermal diffusivity exceeds the mass diffusivity, the $Ka^{1/3}$ scaling is lost.

\subsection{Parameterization for turbulent combustion speed \label{subsec:subsection}} \addvspace{10pt}

We now focus on the search for a suitable scaling function for the turbulent combustion speed, capable of fitting the experimental results seen thus far. We begin by analyzing the sensitivity of  the turbulent combustion speed to the Karlovitz number.
Concerning the logarithmic plot in fig.~\ref{fig:loglog_STSL}, 
the thermodiffusively unstable regime ($Le<Le_0$) is represented by dashed segments whereas the
stable regime ($Le>Le_0$) with continuous segments, both representing power laws found by   
by linear regressions of the kind of $y =mx +p$. Figure \ref{fit_ST_exponent} shows that both the exponents $m$ and the constants $p$
can be well approximated by linear functions of the Karlovitz number.  
By introducing the {\it ansatz} that such parameters are functions of the Karlovitz, $Ka$ number alone,
one may represent the turbulent combustion speed as:

\begin{figure}[h!]
\centering
\includegraphics[width=0.5\textwidth]{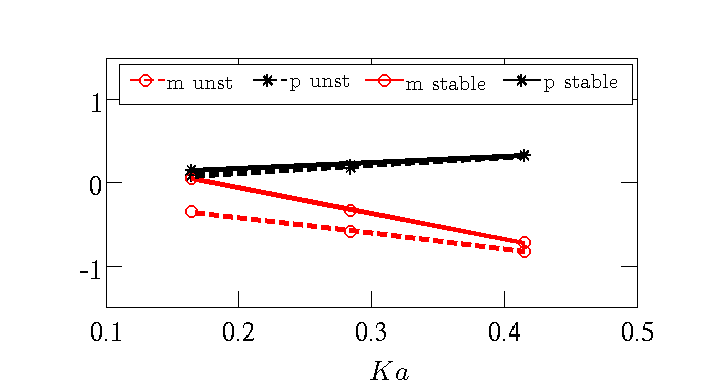}
\caption{Fitting parameter of turbulent combustion speed power laws of figure \ref{fig:loglog_STSL}.
The fitting function is of the kind of $y = mx+p$. Labels \textit{unst} and \textit{stable} refer to regressions for $Le_{eff} < Le_0$ and $Le_{eff} < Le_0$, respectively.}
\label{fit_ST_exponent}
\end{figure}

\begin{equation}
\frac{S_T}{S_L^0} \left( Ka, Le \right) =
A \left( Ka \right) Le^{B \left( Ka \right) } ~.
\label{exp_f}
\end{equation}
%
In logarithmic form, this reads:
\begin{equation}
Log_{10} \left( \frac{S_T}{S_L^0}  \right)=
\underbrace{Log_{10} \left( A  \right)}_ {C}+ B  Log_{10} \left( Le \right) ~,
\end{equation}
 with 
 dependencies from Karlovitz and Lewis removed for conciseness and 
 $10^C =A$. Plugging such results into equation \eqref{exp_f} with
$C = \zeta_p(Ka)$ and $B = \zeta_m(Ka)$, we obtain
 
\begin{equation}
 \frac{S_T}{S_L^0} \left( Ka, Le \right) =   10^{\zeta_p(Ka)}~Le^{\zeta_m(Ka)}~,
\label{ansatz_1}
\end{equation}
with different values for $\zeta_p$ and $\zeta_m$, in the stable (subscript '$s$') and unstable ('$u$') region.
Linear regressions of the data in figure \ref{fit_ST_exponent} yield exponents $\zeta_p(Ka)$ and $\zeta_m(Ka)$. More precisely  
when $Le < Le_0$
\begin{equation}
\begin{split}
 \zeta_{pu}(Ka)= -0.08+0.96 \cdot Ka \\
\zeta_{mu}(Ka) = -0.04 -1.90 \cdot Ka
\end{split}
\label{ka_unst}
\end{equation}
and when $Le > Le_0$
\begin{equation}
\begin{split}
 \zeta_{ps}(Ka) = 0.03 +0.72 \cdot Ka \\
\zeta_{ms}(Ka)= 0.55-3.08 \cdot Ka
\end{split}
\label{ka_st}
\end{equation}

\begin{figure}[h!]
\centering
\includegraphics[width=0.5\textwidth]{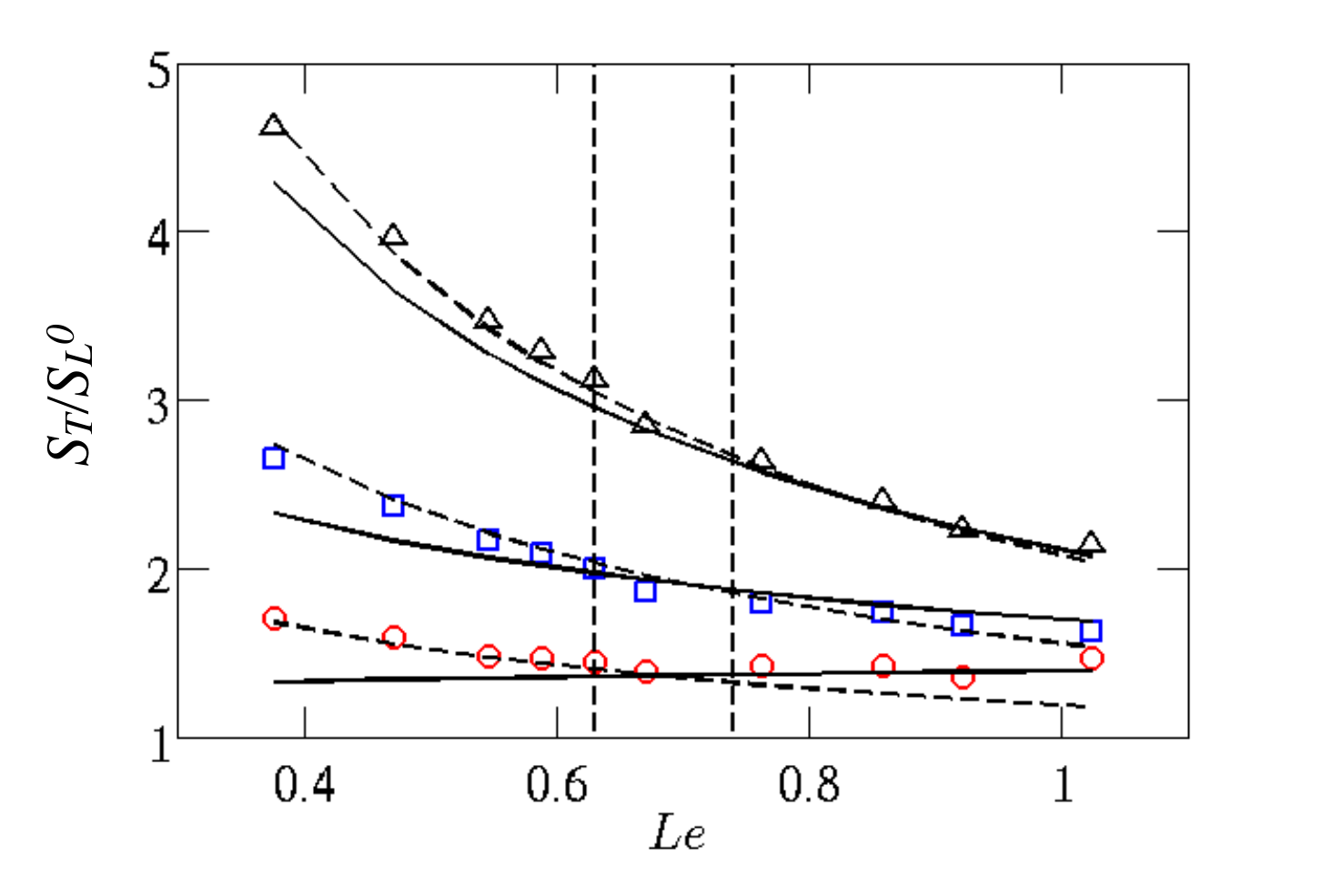}
\caption{turbulent combustion speed: experiments vs fitting functions. Dashed lines uses parameters defined in equation \eqref{ka_unst}. Continuous lines refer to those in equation \eqref{ka_st}.}
\label{mod_Ka_Le}
\end{figure}

The system of equations \eqref{ansatz_1},\eqref{ka_unst} and \eqref{ka_st} can be used to model
the turbulent combustion speed at varying Lewis and Karlovitz numbers as shown in figure \ref{mod_Ka_Le}, where all the experimental data, belonging to the stable/unstable regions, are well fitted and the critical Lewis number is found at the intersection of the two scaling laws. 
At high $Ka$ (black triangles) the two curves are almost superimposed, thus supporting once again
the hypothesis that turbulence acts to smooth the transition between thermal-diffusive stable/unstable regimes.
Indeed, this is also seen by the converging trend of the two angular coefficients $m$, shown in figure \ref{fit_ST_exponent} and representing the exponent of the scaling law of equation \ref{exp_f}.

\begin{figure}
\centering
\includegraphics[width=0.5\textwidth]{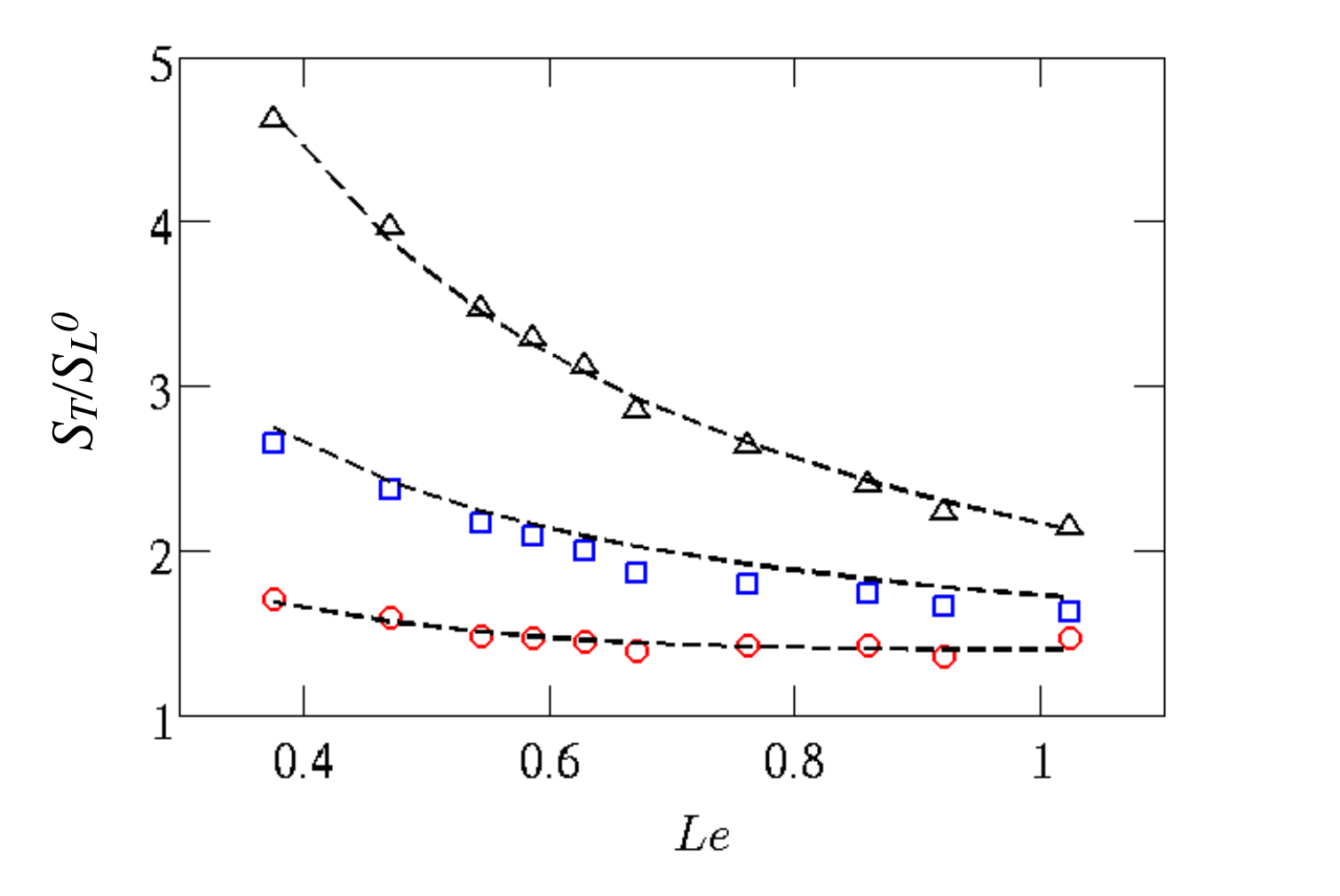}
\caption{Turbulent combustion speed: experiments vs unified fitting functions of equation \ref{attmept}.}
\label{comp_fit}
\end{figure}

An attempt to find an unifying function to describe the turbulent combustion speed being able to model the cross-over through the critical Lewis number $Le_0$ can be made by introducing  
\begin{equation}
\begin{split}
& \frac{S_T}{S_L^0} \left( Ka, Le) \right) = \frac{10^{\zeta_{pu}} \cdot Le^{\zeta_{mu}}}
{\left( 1+  10^{\beta} \cdot (Le/Le_0)^{h} \right)^{\gamma/h}} \\
& \beta = \zeta_{ps} - \zeta_{pu} \\
& \gamma = \zeta_{mu} -\zeta_{ms}
\end{split}
    \label{attmept}
\end{equation}
When $Le<Le_0$ the model behaves as:
\begin{equation}
\frac{S_T}{S_L^0} \left( Ka, Le \right) \propto {10^{\zeta_{pu}} \cdot Le^{\zeta_{mu}}}
    \label{attmept_u}
\end{equation}
and when $Le>Le_0$ 
\begin{equation}
\frac{S_T}{S_L^0} \left( Ka, Le \right) \propto 10^{\zeta_{ps}} \cdot Le^{\zeta_{ms}}~.
    \label{attmept_s}
\end{equation}
In this case, the constant $h$ is used to sharpen the transition across 
$Le_0$ (chosen as $Le_0 = 0.67$) and a suitable value is taken equal to $h = 5$. 
The result is plotted in figure \ref{comp_fit}.


%
\section{Conclusion} \addvspace{10pt}
In this work, the turbulent combustion speed $S_T$ is measured for a Bunsen flame at different turbulence intensities and a wide range of Lewis numbers. Results show that $S_T$ increases as the Lewis number decreases while sensitivity to the turbulent velocity fluctuations is observed, especially when the fuel blend is characterized by low $Le$.
The increase of turbulent combustion speed is shown to be caused by an enhancement of the reactivity of the flame front, rather than by the increase of the normalized turbulent flame area. 
Finally, a parameterization of the experimental results is proposed, based on a functional form depending on both Karlovitz and Lewis numbers, which is also capable of capturing the transition represented by the critical Lewis number $Le_0$.

In the future, while the range of Lewis number explored in this work is sufficient, more work is needed to span wider variations of the Karlovitz as well as the Reynolds numbers, in order to assess the validity of the results obtained or establish a more complex behavior of the functional fit proposed for the experimental results.

\acknowledgement{Declaration of competing interest} \addvspace{10pt}
The authors declare that they have no known competing financial interests or personal relationships that could have appeared to influence the work reported in this paper.

\acknowledgement{Acknowledgments} \addvspace{10pt}

P.E.L acknowledges the support of Sapienza University for the early-stage researchers' funding for the projects “A pragmatic support to the hydrogen economy: data-driven modeling of high-pressure combustion for propulsion and power” and "Combustion under extreme thermodynamic conditions for green propulsion and power". F.D. and F. C. acknowledge the support of Baker-Hughes and Lazio Region. This work has been also supported by ICSC (Centro Nazionale di Ricerca in High-Performance Computing, Big Data and Quantum Computing) funded by the European Union – NextGenerationEU.

 \footnotesize
 \baselineskip 9pt


\bibliographystyle{pci}
\bibliography{PCI_LaTeX}


\newpage

\small
\baselineskip 10pt



\end{document}